\documentclass{article}
\usepackage[utf8]{inputenc}
\usepackage{authblk}
\usepackage{setspace}
\usepackage[margin=1.25in]{geometry}
\usepackage{graphicx}
\graphicspath{ {./figures/} }
\usepackage{subcaption}
\usepackage{amsmath}
\usepackage{lineno}
\usepackage{xcolor}

\usepackage[T1]{fontenc}
\usepackage{textcomp}
\usepackage{lmodern}
\usepackage{newunicodechar}
\newunicodechar{̈}{\"{}}

\RequirePackage[uncertainty-mode=separate, 
    separate-uncertainty-units=single,
    exponent-product=\ensuremath{
\cdot}]{siunitx}

\RequirePackage{graphicx, bm, nicefrac}

\usepackage[style=nejm, 
citestyle=numeric-comp,
sorting=none]{biblatex}
\title{Laser-induced transient opacity in helium nanodroplets probed by single-shot coherent diffraction}

\author[1,2*]{Julian C. Schäfer-Zimmermann}
\author[3]{Tom von Scheven}
\author[1]{Alessandro Colombo}
\author[1,2]{Katharina Kolatzki}
\author[3]{Björn Kruse}
\author[4]{Bruno Langbehn}
\author[4]{Thomas Möller}
\author[2]{Nils Monserud}
\author[1,2]{Mario Sauppe}
\author[2]{Bernd Schütte}
\author[1,2]{Björn Senfftleben}
\author[2]{Rico Mayro P. Tanyag}
\author[4,5]{Anatoli Ulmer}
\author[2,3]{Thomas Fennel}
\author[2]{Marc J.J. Vrakking}
\author[2]{Arnaud Rouzée}
\author[1,2*]{Daniela Rupp}

\affil[1]{Laboratory for Solid State Physics, ETH Zurich, 8092 Zurich, Switzerland}
\affil[2]{Max Born Institute for Nonlinear Optics and Short Pulse Spectroscopy, 12489 Berlin, Germany}
\affil[3]{Institute of Physics, University of Rostock, 18051 Rostock, Germany}
\affil[4]{Institute of Optics and Atomic Physics, Technical University of Berlin, 10623 Berlin, Germany}
\affil[5]{Department of Physics, University of Hamburg, 22761 Hamburg, Germany}
\affil[*]{Address correspondence to: j.schaeferzimmermann@protonmail.com, ruppda@phys.ethz.ch}

\begin{document}
\maketitle


\begin{abstract}
Single-shot coherent diffractive imaging (CDI) with intense short-wavelength light pulses enables the structural characterization of individual nanoparticles in free flight with high spatial and temporal resolution. Conventional CDI assumes that the target object exhibits a linear scattering response and static electronic properties. Here, we extend this approach to investigate transient laser-driven modifications of the electronic structure in individual nanoparticles, imprinted in their time-resolved diffraction patterns. In the presence of a near-infrared laser pulse, we observe a pronounced reduction in the diffraction signal from helium nanodroplets when probed with ultrashort extreme ultraviolet (XUV) pulses. This effect is attributed to a light-field-induced modification of the electronic structure of the droplets, which substantially increases their XUV absorption.
Our results demonstrate  the possibility to capture ultrafast light-driven electron dynamics in nanoscale systems with single-particle diffraction. This opens a pathway towards the  spatiotemporal tracking of reversible changes in the electronic properties of nanoscale structures with potential applications in ultrafast X-ray optics, materials science, and all-optical signal processing.
\end{abstract}


\section{Introduction}
Single-shot Coherent Diffraction Imaging (CDI) with intense X-ray or extreme ultraviolet (XUV) pulses has revolutionized structural analysis at the nanoscale, enabling the imaging of fragile and non-reproducible nanoparticles in free flight \cite{Bogan2008,Sun2022,Aquila2015}. This includes single viruses and bacteria \cite{Seibert2011,Ekeberg2015,vanDerSchot2015,Reddy2017}, aerosols \cite{Loh2012,Colombo2025}, atomic clusters \cite{Bostedt2010, Rupp2012, Rupp2014, Barke2015, Colombo2023}, and helium nanodroplets \cite{Gomez2014, Tanyag2015, Jones2016, Bernando2017, Rupp2017, Langbehn2018, Oconnell2020, Verma2020, Feinberg2021, Feinberg2022, Ulmer2023}. Using pump-probe techniques, CDI can even be employed to track extremely fast structural dynamics  in isolated nanoparticles  following pulsed laser excitation, such as ultrafast melting,  breathing and boiling \cite{Shin2023, Hoeing2023, Dold2025} or Coulomb driven explosion \cite{Gorkhover2016, Flueckiger2016, Bacellar2022, Langbehn2022, Peltz2022}.

Traditionally, in both static and time-resolved single-shot single-particle CDI, we operate under the so-called "diffraction before destruction" assumption \cite{Chapman2014}. The analysis of diffraction patterns is carried out based on the approximation that throughout the illumination with the X-ray pulse the scattering density remains constant. Relevant X-ray-induced structural damage occurs only after the pulse is over. Any electron dynamics during the diffraction process such as the unavoidable ionization by the short-wavelength photons is usually also neglected, which puts a clear limitation to the accurate interpretation of diffraction patterns. Seen from a different perspective, the fact that diffraction encodes both the target’s geometry and its optical response can also be treated as a feature. In this sense, single-shot single-particle CDI inherently offers a way to monitor ultrafast modifications of a nanoparticle’s electronic structure, such as ionization-driven changes \cite{Bostedt2012, Ho2020, Rupp2020, Kuschel2025}, in a near-background-free manner. Until recently, resolving such processes in time was not feasible. With the recent advent of intense attosecond pulses from high-harmonic generation (HHG) sources \cite{Midorikawa2022,Kretschmar2024,Martin2025} and short-wavelength free-electron lasers (X-FELs) \cite{Duris2020, Maroju2021, Nolting2023,Yan2024,Franz2024,Guo2024,Prat2025} attosecond CDI has become possible, thus enabling diffraction snapshots of the system under investigation on electronic timescales \cite{Kuschel2025}. This has inspired theoretical work, with early studies suggesting that CDI is able to capture intricate ultrafast electronic changes \cite{Dixit2013, Popova2018, Kruse2020, Popova2024, Venkatesh2025}. Furthermore, methods for handling broadband attosecond diffraction have been proposed \cite{Abbey2011, Witte2014, Sander2015, Malm2020, Huijts2020}.

The ultrafast nonlinear electronic response of laser-irradiated matter is of fundamental interest for a broad range of scientific disciplines. In atomic systems, understanding strong-field interactions has enabled the generation of attosecond pulses \cite{Corkum2007, Krausz2016}, marking the beginning of attosecond science \cite{Krausz2014}. This has provided time-resolved access to fundamental electron-driven phenomena \cite{Krausz2009, Eckle2008, Sansone2010, Smirnova2009}, as recognized by the 2023 Nobel Prize in Physics \cite{LHuiller2023}. Beyond fundamental studies, nanoscale strong-field effects have practical applications in all-optical signal processing \cite{Stockman2011, Ciappina2017}, for the development of switchable dielectric nanostructures for next-generation electronics \cite{Hassan2024,Heide2024} and in quantum information science \cite{Bhattacharya2023}. While attosecond transient absorption spectroscopy (ATAS) \cite{Gallmann2013, Goulielmakis2010, Holler2011, Chini2012, Chen2012, Ott2014, Beck2014, Reduzzi2015, Liao2016, Lucchini2016, Jordan2020, Niedermayr2022,Cavaletto2025} is a well-established method for tracking strong-field laser-induced dynamics, a spatially resolved single-particle imaging technique with similar sensitivity to transient electronic structures could provide fundamentally new insights.

In this work, we drive electron dynamics in an isolated nanoscale target using a moderately intense near-infrared (NIR) laser pulse and probe the response via time-delayed single-shot diffraction with near-resonant XUV pulses from an intense HHG source. The NIR pulse alone passes the helium nanodroplet without ionizing it, as its intensity is far below the ionization threshold. However, when both NIR and XUV pulses are present and temporally overlapped, we observe a pronounced reduction in brightness in the XUV diffraction image. We attribute this effect to a reversible NIR-induced modification of the electronic structure of the helium droplet. Our interpretation is supported by simulations, which incorporate an NIR-induced effective refractive index derived from time-dependent Schrödinger equation (TDSE) calculations. They indicate that the dominant NIR-induced effect is a transient increase in the imaginary part of the refractive index at the photon energy of the strongest XUV harmonic owing to a light-field-induced shift of an absorption band in liquid helium toward this energy, thus turning the droplets from almost transparent but strongly refracting to fully opaque.

Our results demonstrate that time-resolved single-shot diffraction is able to track reversible optically controlled electronic structure modifications in isolated nanoscale systems. With the emergence of high-intensity attosecond (soft) X-ray pulses \cite{Midorikawa2022,,Kretschmar2024,Martin2025,Duris2020,Maroju2021,Nolting2023,Yan2024,Franz2024,Guo2024,Prat2025}, future experiments will allow to visualize the spatio-temporal evolution of laser-driven electronic modifications, such as field-induced level shifts \cite{Autler1955, Delone1999}, the formation of light-induced states \cite{Chen2013}, or ultrafast band dynamics such as the dynamical Franz-Keldysh effect \cite{Lucchini2016,Otobe2017,Dolso2025} with nanometer spatial resolution and sub-optical-cycle time resolution.

\begin{figure}[t!]
    \includegraphics[width=1\linewidth]{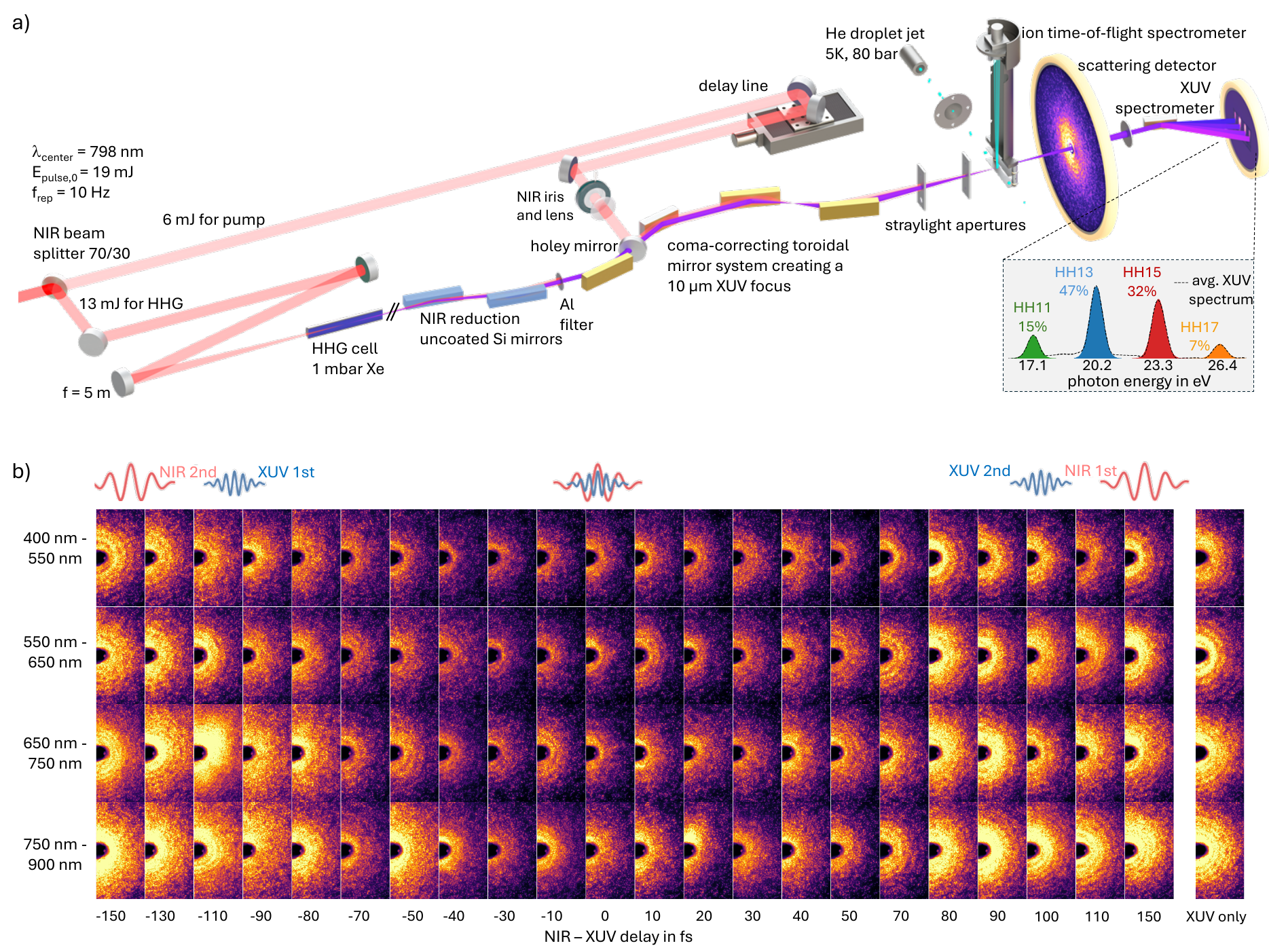}
    \caption{{\bf a)} Sketch of the experimental setup for time-resolved single-shot coherent diffraction imaging of helium nanodroplets with an intense HHG source. The inset shows the average contribution of individual harmonics to the total XUV intensity. {\bf b)}  The brightest size-selected diffraction image for each of four size ranges in the center of the observed size distribution as a function of XUV-NIR time delay. Delays with low statistics were omitted. The rightmost column shows the brightest size selected image for the four size ranges from an XUV-only data set of comparable size to each delay measurement (color scale see Fig. \ref{fig:dip}a, full-resolution data set see \emph{Data Availability}).}
    \label{fig:exp_setup_v4}
\end{figure}

\section{Experiment}

The experimental setup for time-resolved single-shot coherent diffraction of helium nanodroplets is illustrated in Fig. \ref{fig:exp_setup_v4} a). An intense HHG source \cite{Schutte2016, Rupp2017} was driven by NIR pulses from a Ti:Sa laser amplifier ($E_{\text{pulse}}$=\,\SI{13}{\milli\joule}, $f_{\text{rep}}$=\,\SI{10}{\hertz}, $\lambda_{\text{center}}=\,$\SI{798}{\nano\meter}, $t_{\text{NIR,FWHM}}=$\,\SI{40}{\femto\second} at the output, increasing to \SI{60}{\femto\second} after \SI{20}{m} propagation in air). Using a loose-focusing geometry, about \SI{2}{\micro\joule} of XUV light (pulse energy measured with a calibrated XUV photodiode, pulse duration estimated to $t_{\text{XUV,FWHM}}$=\,\SI{25}{\femto\second}) were generated in a \SI{10}{\centi\meter} long xenon-filled cell ($p_{Xe}$= \SI{1}{\milli\bar}). The residual NIR light was reduced by gracing-incidence reflection of two uncoated Si blocks and subsequently further suppressed by a \SI{100}{\nano\meter} thick Al filter foil. The XUV beam was focused by a system of toroidal mirrors \cite{Frassetto2014} to a \SI{10}{\micro\meter} spot, reaching a focal intensity of \SI{1.3}{\tera\watt\per\square\centi\meter}, sufficient for generating bright single-particle diffraction patterns \cite{Rupp2017}.

At the focus, a jet of helium nanodroplets perpendicularly intersected the XUV beam. The droplets were produced by expanding liquid helium at  \SI{5}{\kelvin} and \SI{80}{\bar} into vacuum through a pulsed \SI{100}{\micro\meter} trumpet nozzle. These conditions have been characterized previously \cite{Langbehn2018} to produce a mean droplet size of $\text{<}N\text{>}=$ \num{6e9} atoms ($R\approx\,$\SI{400}{\nano\meter}). A conical skimmer with 0.5 mm diameter further collimated the droplet jet to ensure single-droplet interactions in the XUV focus. The diffracted light from each intercepted droplet was converted into optical photons by a large-area MCP/phosphor stack reaching scattering angles up to $45^{\circ}$ and recorded by an out-of-vacuum camera \cite{Bostedt2010}. The unscattered part of the XUV beam was guided through a hole in the MCP detector onto a grating to measure the XUV spectrum on a shot-to-shot basis. The average spectrum (see inset in Fig. \ref{fig:exp_setup_v4}$\,$a)) shows a dominant contribution from the $\num{13}\text{th}$ and $\num{15}\text{th}$ harmonics at \SI{20.2}{} and \SI{23.3}{\electronvolt}, respectively.

For time-resolved measurements, a second NIR pulse from the same Ti:sapphire laser amplifier was introduced into the beam path with a controlled time delay relative to the XUV pulse using a holey mirror. Due to its transport through \SI{30}{m} air and its passage over several optical elements, the temporal width of this NIR pulse was broadened to $t_{\text{FWHM, NIR}}\approx\,$\SI{80}{\femto\second}. An out-of-vacuum delay-stage allowed for scanning XUV-NIR delay. The timing uncertainty  between the pulses was estimated to be \SI{14}{\femto\second} r.m.s. (see Supplementary Note 1). Spatial overlap of the NIR and XUV foci was obtained by adjusting a lens outside the vacuum chamber. To ensure that the NIR pulse alone did not ionize the helium droplets, its intensity was reduced by closing an iris until no ion signal was detected in simultaneously recorded time-of-flight (tof) spectra. The resulting pulse energy of \SI{0.40}{\milli\joule} in a focal spot of \SI{250}{\micro\meter} in diameter corresponds to an intensity of \SI{20}{\tera\watt\per\square\centi\meter}. The much larger size of the NIR focus allowed for a rather stable spatial overlap of both foci. To further minimize the impact of experimental drifts, short acquisitions of 3000 shots were taken, scanning XUV-NIR delays in random order, and overlap checks were performed regularly (see also Supplementary Note 2).

\section{Results}

In a preprocessing step,  single-droplet "hits" (approximately 16000 images, \~5\% of recorded patterns)  were selected, based on the brightness of the diffraction images and the simultaneously recorded helium ion yield (see Supplementary Note 2 for details of the hit selection).

 By obtaining the radial profiles of each pattern and Fourier transforming their derivative, we can assign an approximate droplet size to the majority of diffraction patterns (see supplementary note 8 for details on the procedure and supplementary note 9 for the measured size distribution). Conclusions from such a size analysis have to be drawn very carefully for several reasons. First, the measured size distribution is generally strongly intensity dependent. At a given incoming focal power density small droplets do not cast sufficient signal to be detected at all. From a certain size on, only droplets in the center of the focus are detected. Very large droplets can be imaged in an extended area of the focal volume. This strongly biases any size distribution measured with single-particle CDI towards large sizes and creates a shift in the size distribution when the intensity changes. The use of a laboratory-based HHG source in our work reduces the available intensities by orders of magnitude as compared to short-wavelength FELs \cite{Langbehn2018}, being reflected in an observed size distribution shifted towards larger values in our work.
The sizing in the delay region of decreased intensity is additionally affected by this intensity dependence. Furthermore, the relation between the extracted spatial frequency and a droplet size is depending on the optical properties, resulting in uncertainties of up to 20\% (see supplementary note 8). 

By binning the size selected patterns into four broad size ranges of at least 100 nm width in the center region of the observed size distribution, appropriate comparisons can still be conducted. Fig. \ref{fig:exp_setup_v4} b) displays the brightest diffraction image for each time delay and size range.  A clear brightness reduction can be observed in the region of temporal overlap of the pump and probe pulses  for all size ranges. 

\begin{figure}[t]
        \includegraphics[width=0.4\linewidth]{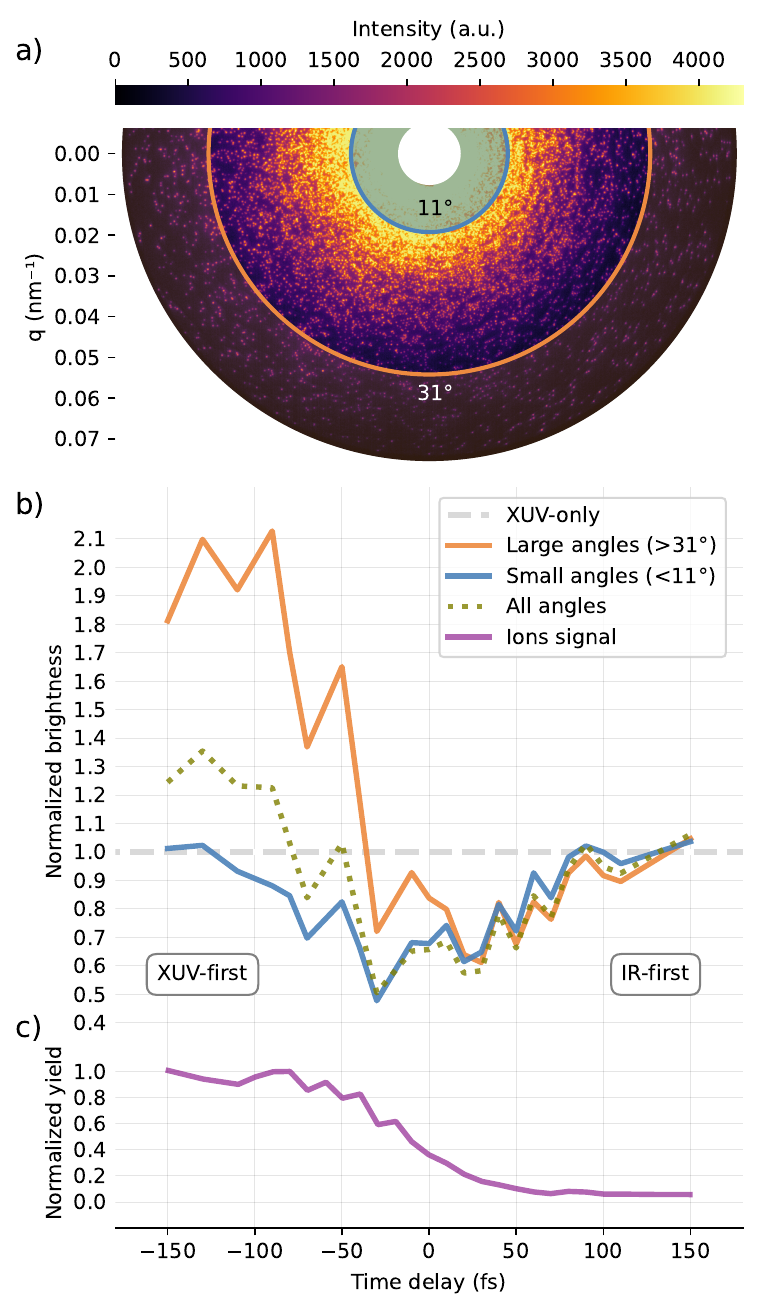}
        \caption{\label{fig:results_decrease_tof_det_signal}
            {\bf a)} Exemplary diffraction pattern with indicated inner detector area, in blue up to \SI{11}{\degree} scattering angle and outer detector area in orange, from \SI{31}{\degree} up to the edge of the detector at \SI{45}{\degree}.
            {\bf b)} Delay-dependent integrated detector signal for all hits, restricted to different areas of the diffraction detector (full detector signal as green dotted line, inner detector area as blue solid line, outer detector area as orange solid line). All curves are normalized to the respective XUV only signal. {\bf c)} Delay-dependent integrated helium ion yield, normalized to its maximal value.  
        }\label{fig:dip}
\end{figure}%

For the subsequent analysis, we focus on the integrated detector signal  as a function of XUV-NIR time delay. The full detector signal  (center hole and detector areas of decreased efficiency masked out, see also Supplementary Note 4)   of all "hits", normalized to the XUV-only signal, is shown as light-green dotted curve in Fig. \ref{fig:results_decrease_tof_det_signal} b). While a region of decreased intensity is clearly visible around zero time delay, towards negative delays the signal rises above the XUV-only level. This increase can be assigned to  fluorescence radiation which also produces a signal on the MCP detector. At negative delays, i.e. when the XUV pulse precedes the NIR pulse, a process termed "avalanching" \cite{Schutte2016} occurs: The helium droplets are already partially ionized when the NIR arrives, thus allowing the NIR field to efficiently drive and heat droplet-bound quasi-free electrons. This has been shown to result in strong collisional ionization \cite{Schutte2016}. However, collisional excitation and recombination in the nanoplasma will also lead to the formation of excited atoms and ions, which can eventually emit fluorescence light \cite{Schroedter2014, Muller2015}. Unlike the strongly forward-directed diffraction signal, which drops with $q^{-4}$ , fluorescence is emitted isotropically, making it  a potentially very significant  contribution in the outer detector area. While the non-linear response function of the MCP based diffraction detector does not allow for subtracting a constant fluorescence signal, we minimize its influence by restricting the subsequent analysis of the integrated detector signal to small scattering angles up to $11^{\circ}$ (marked blue in Fig. \ref{fig:results_decrease_tof_det_signal} a), where the diffraction dominates by orders of magnitude. 

Accordingly,  comparing the integrated detector signals from small (blue solid line in Fig. \ref{fig:results_decrease_tof_det_signal} b) and large ($>31^{\circ}$, marked orange in Fig. \ref{fig:results_decrease_tof_det_signal} a, orange solid line in Fig. \ref{fig:results_decrease_tof_det_signal} b) scattering angles, normalized to the respective XUV-only signals, we observe two main trends. At small angles where diffraction dominates, the signal shows a pronounced intensity reduction of up to 30\%, which is rather symmetric around zero delay.  In contrast, the detector signal at large angles strongly increases towards negative delays. While at positive delays above 50\,fs, the large angle signal still follows the general "dip" behavior, as soon as the helium ion signal shown in Fig. \ref{fig:results_decrease_tof_det_signal} c, which is a known avalanching signature, rises above its baseline, the high-angle signal diverges. The Pearson correlation between single-shot helium ion and high-angle detector signals yields of 0.94 or higher for all negative delays values, which confirms a common cause of the increase.  

\section{Modeling}

\begin{figure}[t!]
    \includegraphics[width=0.85\linewidth]{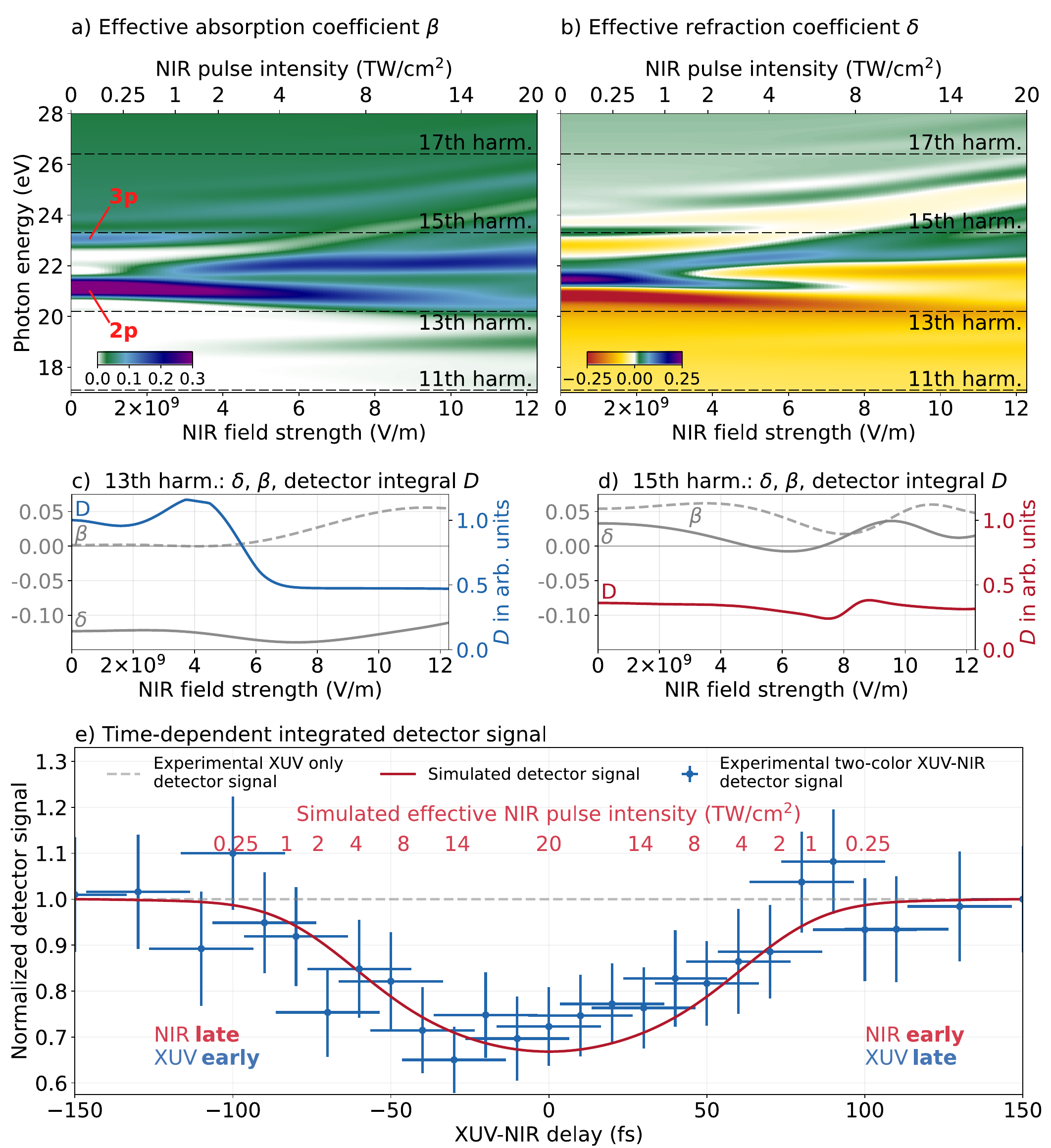}
            \caption{
        {\bf a, b)} Simulated effective refractive index of helium droplets around \SI{20}{\electronvolt} photon energy (absorption coefficient $\beta$, refraction coefficient $\delta$) vs. NIR field strength (bottom axis) or laser intensity (top axis). {\bf c)} Line-outs of the effective refractive indices at the positions of the 13th and {\bf d)} 15th harmonic are given as gray solid ($\delta$) and dashed ($\beta$) lines. In addition, the detector integrals at these energies for an exemplary droplet size of  R = 708 nm  are given in blue and red. The same mask as for the experimental patterns (Supplementary Fig. 7) is applied. {\bf e)} Measured (blue markers) and simulated (red curve) delay-dependent detector signal as a function of the NIR-XUV delay. The effective average NIR intensities during the XUV scattering process are indicated above the curve.}
        \label{fig:results_decrease_csca_det_signal_ref_index_double_column}
\end{figure}

Our model of the delay-dependent diffraction response of laser-dressed helium nanodroplets is leaning on the similarity of the electronic structure of atomic and bulk helium in the absence of an external laser field. The absorption, reflection, and refractive index of liquid helium in the XUV range are documented in the literature \cite{Surko1969, Lucas1983, Joppien1993}. They mimic the equivalent properties of atomic helium, exhibiting dominant resonance features at the $1s^{2}\rightarrow1snp$ (n=2,3,...) transitions, with small blue shifts and broadening of the narrow transitions into bands due to van der Waals interactions in the liquid. Additionally, in nanodroplets, small contributions from optically forbidden transitions appear, such as $1s^{2}\rightarrow1s2s$, as these transitions become allowed at the surface \cite{Joppien1993}. 

 To describe the optical properties of the NIR-laser-dressed Helium droplets, we depart from an effective atomic response as predicted by the numerical solution of the time-dependent Schrödinger equation for isolated helium atoms under an intense NIR dressing field. The resulting effective optical properties in the XUV range are determined by applying a generalized form of the well-known kick-response method \cite{Gaarde2011, Wu2016}. Therefore the dipole response of the NIR-driven atomic system to an additional broadband electromagnetic "kick" is recorded and Fourier transformed to extract the effective dynamic polarizability. The finite width of the optical transitions in liquid helium is introduced by using an artificial decoherence time \cite{Gaarde2011, Chen2012} of \SI{8}{\femto\second} that is implemented by multiplying the dipole moment with a corresponding Hann window function before Fourier transformation. Using the resulting simulated dynamical polarizability and averaging over the carrier envelope phase of the NIR pulse, an effective and intensity-dependent refractive index $n=1-\delta+i\beta$ was derived for bulk helium in the presence of the NIR field via the Clausius-Mossotti relation (for details see Supplementary Note 4.1).

This procedure delivers a physically transparent approximation of the spectrally-resolved effective refractive index with a consistent oscillator strength, while reproducing the width of the 1s2p transition of the literature values for liquid helium at zero field \cite{Lucas1983, Surko1969} (see Supplementary Note 4.4). In the interest of a clear picture of the NIR dressing effect we refrain from further artificial fixes to mimick the blue shift of transitions or the droplet-size effect of optical properties. 

Simulated effective refractive index maps as a function of NIR intensity are presented in Fig. \ref{fig:results_decrease_csca_det_signal_ref_index_double_column} a) and b). The absorption coefficient $\beta$, shown in Fig. \ref{fig:results_decrease_csca_det_signal_ref_index_double_column} a), corresponds to the imaginary part of the refractive index and therefore to how quickly light propagation decays in the droplet. The refraction coefficient $\delta$, displayed in Fig. \ref{fig:results_decrease_csca_det_signal_ref_index_double_column} b), is the deviation of the real part of the refractive index from unity and relates to how strongly light of a certain photon energy is refracted in the droplet. The positions of the main harmonics in the experiment, i.e. harmonics 11 to 17, are marked as dashed black lines. Figs. \ref{fig:results_decrease_csca_det_signal_ref_index_double_column} c) and  d) show line-outs of $\beta$ and $\delta$ at the central energies of the dominant 13th and 15th harmonics, respectively. They are compared to the calculated integrated detector signal $D$ for an average droplet radius of \SI{708}{\nano\meter}, shown in blue and red, providing a rough estimate for the contributions of these harmonics to the observed dynamics. For this purpose, the single-color diffraction patterns of an NIR-dressed helium nanodroplet of this size were calculated at the central wavelengths of the 13th and 15th harmonic using Mie simulations \cite{BohrenHuffman} and integrated over the same area as the experimental data (i.e. up to $11^{\circ}$ scattering angle, areas masked as in Supplementary Fig. 7). A full map of the size-dependent harmonic contributions is given in Supplementary Note 6.

To directly compare the simulation results with our measured signal, we calculated the integrated detector signal at small angles for all harmonics (using the average XUV spectrum shown in the inset of Fig. 1 a)) at the effective dressing laser intensities experienced at each XUV-NIR time delay (Supplementary Note 5.5), accounting also for the nonlinear detector response and the measured size distribution of the helium nanodroplets (see supplementary notes 4 and 9). The final simulated time-dependent detector signal is displayed (red solid line) in Fig. \ref{fig:results_decrease_csca_det_signal_ref_index_double_column} e) alongside the experimental measurements (blue markers), both normalized to the respective XUV-only signal (gray dashed line), showing a remarkable agreement between experiment and calculation.

 The small-angle detector signals from size selected diffraction patterns for the four size ranges as defined in Fig.\,\ref{fig:exp_setup_v4}$\,$b are shown in Fig.\,\ref{fig:size_selected_main}. A tendency of a deepening dip for smaller droplets in the experimental data is observed which is also well reproduced by the simulation.

\section{Discussion and Conclusion}

\begin{figure}[t]
        \includegraphics[width=1\linewidth]{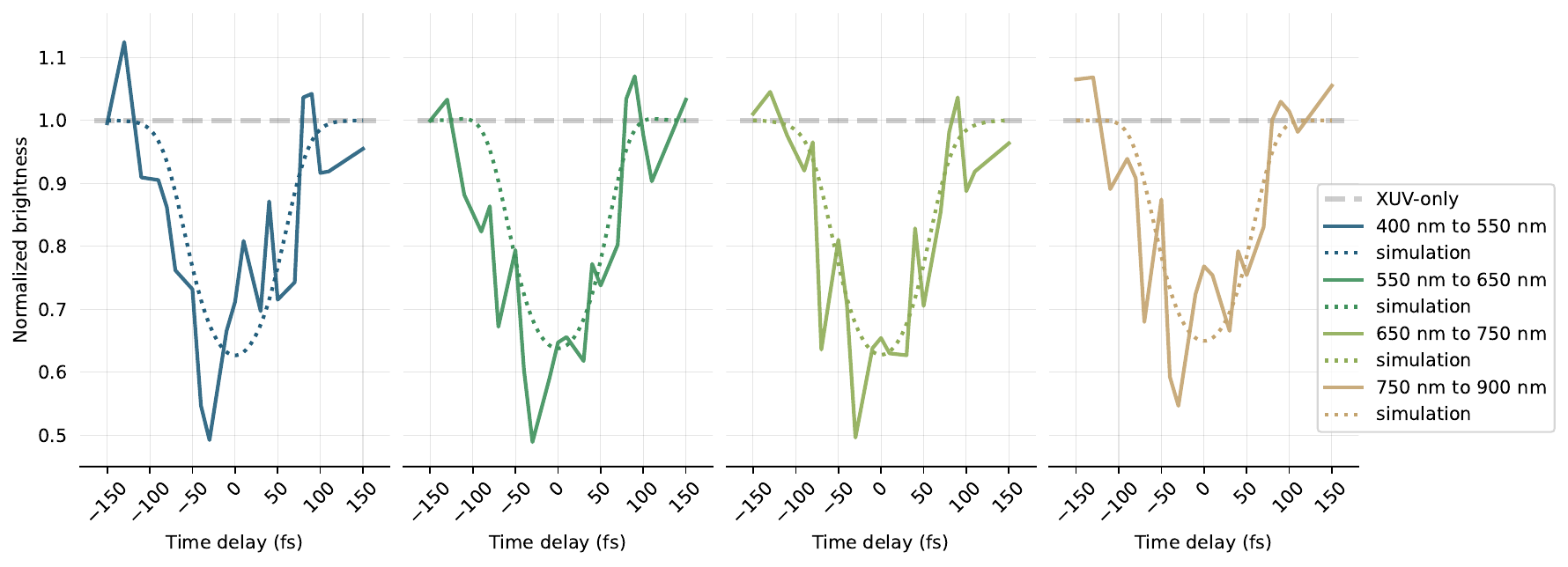}
        \caption{\label{fig:size_selected_main}
             Delay dependent small-angle detector signal of size selected diffraction patterns for four different size ranges. Simulated curves for single sizes are given as dotted lines. 
        }
\end{figure}%

The simulated diffraction signal reproduces well the measured time-dependent low-angle detector signal with a drop of the diffracted intensity of about $30\%$  in the signal from all hits (Fig. \ref{fig:results_decrease_csca_det_signal_ref_index_double_column} e) and even deeper dips for size selected diffraction patterns (Fig.\,\ref{fig:size_selected_main}). 

This is remarkable since our model is expected to
give correct trends, but not quantitative agreement.   For example, the calculations do not account for the known blue shifts of absorption features in the droplet of up to 0.5 eV compared to the atomic transitions \cite{Surko1969, Lucas1983, Joppien1993} (see Supplementary Note 4.4. for further discussion of discrepancies to literature curves). Furthermore, the kick-response approach doesn't include changes in state occupations during the XUV pulse. Also excitation and ionization within the droplet during the leading edge of the pulse, caused by the XUV or the combined XUV and NIR fields, are neglected. These effects will also modify the diffraction response over the course of the pulse, and in particular may cause an asymmetric behavior of the diffraction response in respect to time zero. However, asymmetric features in the experimentally observed intensity decrease remain within the range uncertainty in time delay and statistical fluctuations of the intensity signal. Furthermore,  full simulations using an XUV pulse train at representative fluences indicate that at the current intensities modifications from ionization should indeed be minor (see Supplementary Note 7). Finally, when calculating the diffraction of a later part of the XUV pulse, spectral reshaping \cite{Chen2013laser} of the light propagating through the droplet is not accounted for. 

With these limitations in mind, the  satisfying agreement between calculation and measurement encourages us to  propose  a basic picture of the physical mechanisms underlying the observed reduction in diffraction signal brightness from the model. For this, we  discuss  the evolution of the effective refractive index maps at the photon energies corresponding to the two dominant harmonics, namely the 13th and 15th, in the XUV spectrum. 

At laser intensities below \SI{1}{\tera\watt\per\square\centi\meter}, the calculated absorption map in Fig. \ref{fig:results_decrease_csca_det_signal_ref_index_double_column} a) shows two prominent features related to absorption corresponding to the $1s2p$ (around 21.4 eV) and $1s3p$ (around 23.2 eV) dipole-allowed transitions, similar to the absorption of atomic helium. With increasing NIR intensity, these bands shift in energy due to the AC Stark effect (the $1s2p$ band is red-shifted and the $1s3p$ band blue-shifted, respectively). We note that sub-cycle features from interfering two-photon quantum paths \cite{Chen2013laser} cannot be resolved in our experiment  and excluded by averaging over the carrier envelope phase of the NIR pulse in our model. Around \SI{1}{\tera\watt\per\square\centi\meter}, a new absorption band appears above the $1s2p$ feature which subsequently shifts upwards to around 22 eV. Comparison with previous studies \cite{Wu2016} suggests that this feature corresponds to a superposition of several light-induced states (LIS), namely $2s+$, $3s-$ and $3d-$ (describing the population of $1s2s$ by the absorption of the difference of one XUV and one NIR photon and $1s3s$ and $1s3d$ by the absorption of the sum of one XUV and one NIR photon). In bulk liquid, these transitions appear to merge into one broad band.

Analog assignments can be made for the general features in the $\delta$-map (refraction), displayed in Fig. \ref{fig:results_decrease_csca_det_signal_ref_index_double_column} b). They are connected to the $\beta$-map landscape via the Kramers-Kronig relations, leading to Fano-like structures in the refraction map which are characteristically shifted in energy compared to the corresponding absorption maxima. 

Focusing on specific photon energies, at the energy corresponding to the 13th harmonic (Fig. \ref{fig:results_decrease_csca_det_signal_ref_index_double_column} c)) our model predicts that the absorption coefficient $\beta$ increases significantly in the presence of the dressing laser field - by an order of magnitude compared to the field-free case - reaching a value above 0.05 at intensities above \SI{10}{\tera\watt\per\square\centi\meter}. This increase is attributed to the progressive shift of the $1s2p$ absorption band into this energy range as the intensity rises. In contrast, the refraction coefficient $\delta$ remains at a nearly constant, high value across all tested intensities at this photon energy, varying only slightly between -0.1 and -0.14. The droplets therefore remain strongly refracting at this photon energy, regardless of the NIR intensity.

At the photon energy corresponding to the 15th harmonic, a different behavior is observed. In the field-free case, the absorption coefficient $\beta$ is relatively high (0.05) due to the proximity of the $1s3p$ absorption band of helium. As the intensity increases, $\beta$ initially decreases, reaching a minimum of 0.02 at \SI{9}{\tera\watt\per\square\centi\meter}, as the $1s3p$ band shifts out of this spectral region. At even higher NIR intensities, the absorption increases again due to the emergence of several light-induced states, with a corresponding absorption shoulder that gradually enters the spectral region corresponding to the 15th harmonic. In contrast to the 13th harmonic, the absolute values of the refraction coefficient $\delta$ at the position of the 15th harmonic are significantly smaller across all NIR intensities, approaching and crossing zero twice in the range between 4 and \SI{8}{\tera\watt\per\square\centi\meter}.

To understand how the evolving optical properties translate into the observed changes in the diffraction response, it is useful to consider several limiting cases for scattering from a spherical particle (see also Supplementary Note 7 for details and visualizations).

When absorption is high (case 1), the particle behaves effectively as an opaque aperture - a condition that depends on the particle size (see Supp. Fig. 11). In this regime, the scattering becomes largely insensitive to changes in the refractive index $\delta$, since light cannot penetrate the object sufficiently for refraction to influence the diffraction pattern. Only when absorption is low can light propagate through the particle, allowing variations in $\delta$ to significantly affect scattering. In the case of low absorption but high refraction (case 2) - the equivalent of a glass sphere in the visible range - the particle exhibits a particularly strong diffraction response. Under these conditions, the scattering intensity can exceed that of an opaque sphere of the same size by up to a factor of four (see Supplementary Note 7). Finally, when both absorption and refraction are low (case 3), the scattering strength is minimal, resulting in a weak diffraction signal. This scenario is typically for the X-ray spectral region.

The calculated detector integral $D$ at the photon energy of the 13th harmonic (blue solid curve in Fig. \ref{fig:results_decrease_csca_det_signal_ref_index_double_column} c)) shows that the droplets exhibit strong scattering at low NIR intensities, thereby dominating the overall diffraction pattern \cite{Rupp2017}. This behavior corresponds to the regime of a "glass sphere", i.e. low absorption and high refraction (case 2), where the particle produces an enhanced diffraction response. However, as the NIR intensity increases, the absorption coefficient $\beta$ rises above 0.02 at around \SI{5}{\tera\watt\per\square\centi\meter}, shifting the system towards the high-absorption regime. In this regime, the particle begins to act more like an opaque aperture (case 1) and further changes in the refraction index have little effect on the scattering. As a result, the contribution of the 13th harmonic to the diffraction pattern drops by approximately a factor of two and then levels off.

In contrast, at the photon energy of the 15th harmonic, the droplet exhibits low refraction values ($\delta$) across all tested NIR intensities (gray solid curve in Fig. \ref{fig:results_decrease_csca_det_signal_ref_index_double_column} d)). As a result, the absorption coefficient $\beta$ becomes the dominant factor influencing the integrated detector signal (red solid curve) at this wavelength. At low intensities, the high absorption ($\beta>$ 0.05) places the system into case 1, where the particle effectively behaves like an opaque aperture. In this regime, changes in $\delta$ have little effect, and the contribution of the 15th harmonic to the overall scattering intensity remains low. As the intensity increases, $\beta$ decreases while $\delta$ at first remains close to zero, driving the system towards case 3, where both absorption and refraction are low. Then, a moderate increase in $\delta$ around \SI{5}{\tera\watt\per\square\centi\meter} allows a stronger diffraction response, and consequently, the scattering signal rises and eventually surpasses its initial value at this intensity before it drops back towards the previous level when $\beta$ rises again. Throughout the tested intensities, the 15th harmonic consistently contributes less to the total diffracted intensity than the 13th harmonic, as evidenced by the direct comparison of their respective contribution to the integrated detector signal (red and blue curves in Fig. 3 c) and d)).

We therefore attribute the pronounced reduction in scattering intensity observed experimentally in the presence of the dressing laser field to the significant increase in the absorption coefficient at the photon energy corresponding to the 13th harmonic. This increase is caused by the NIR-induced AC-Stark shift of the $1s2p$ absorption band, and is therefore directly related to the change of the electronic structure of the droplet in the dressed laser field.

\section{Summary and outlook}

In this work, we demonstrated a  coherent diffraction based experimental approach for characterizing the nonlinear transient electronic response of free-flying nano-samples to laser fields. We observed that during irradiation with a non-ionizing NIR pulse, isolated helium nanodroplets exhibit a substantial decrease in XUV scattering strength at photon energies in the vicinity of \SI{20}{\electronvolt}. We propose an explanation for our findings based on laser dressing of the helium droplets' electronic structure, leading to AC Stark shifts of the atomic-like absorption bands and the occurrence of light induced state-type structures. The observed time-dependent drop of the diffraction signal of about \SI{30}{\percent} on a femtosecond timescale is reproduced by Mie simulations using effective, NIR laser intensity-dependent refractive indices that were computed from \emph{ab initio} TDSE calculations fitted to liquid helium data. In this work, the dominating effect of the NIR field to the diffraction signal from the helium droplets is an increasing absorption at the photon energy of the strongest harmonic, where at zero field, the droplet exhibits enhanced scattering due to strong refraction and weak absorption. However, conceptually both directions are feasible, transiently switching the diffraction response "on" and "off" with a laser, by choosing photon energy and material properties accordingly.

Our additional sizing analysis and the consistent agreement of size selected data with the model allow an outlook towards investigations becoming possible in the future. With an improved setup that also measures the spectra of incoming and diffracted light for each event and the use of a detector system with a linear response function, full retreival of spatial properties and a direct extraction of optical properties, both real and imaginary part, by multicolor-Mie fitting of single-shot diffraction patterns will become feasible.

On the simulation side, our approach can be refined in future work via an improved description that properly includes atomic interactions in the superfluid liquid and resolves bulk and surface contributions to the response in order to retrieve quantitative predictions for transient refractive index dynamics.

Our observations highlight the sensitivity of single-particle CDI to ultrafast changes in the electronic structure of nanoparticles. In this regard, our work establishes CDI as a method sensitive to transient laser-induced electron dynamics in matter that is generally also able to provide spatio-temporal resolution, ultimately down to tens of nanometers and hundreds of attoseconds or less. 

Although short-wavelength free-electron lasers (FELs) consistently deliver well-controlled femtosecond XUV and X-ray pulses for more than 20 years now \cite{Bostedt2009,Allaria2012,Bostedt2016}, only recently extremely intense, few-femtosecond and sub-femtosecond pulses have become available at FELs \cite{Duris2020, Maroju2021, Nolting2023,Yan2024,Franz2024,Guo2024,Prat2025} as well as at high-harmonic generation sources \cite{Midorikawa2022,Kretschmar2024,Martin2025,Rupp2017}. This development will enable further studies to clarify whether other nonlinear effects known from atomic targets, such as the formation of transient multi-photon light-induced states \cite{Chen2012,Wu2016}, can be observed in nanoparticles, how they depend on size and wavelength, and how these phenomena connect to ultrafast absorption effects observed in solids \cite{Lucchini2016}. Theoretical predictions \cite{Kruse2020} of spatio-temporal excitation waves and Rabi-cycling in strongly irradiated nanoparticles leading to significant deviations of the expected diffraction images from the linear response regime further promise that ultrafast CDI will give access to novel quantum phenomena in spatially extended systems. The control of electronic properties of nanoscale matter with light is also a relevant topical issue towards ultimate-speed nano-optoelectronics \cite{Krausz2014,Stockman2011,Heide2024}. Thus, in combination with the rapidly evolving theoretical treatment of attosecond CDI \cite{Rana2020}, our work introduces a powerful new approach for analyzing the nonlinear optical response of arbitrary targets on the nanoscale with potential impact on a broad field of research from ultrafast material science to nonlinear XUV optics and all-optical processing.

\section*{Acknowledgment}
Excellent support has been provided by the MBI technicians and workshops, in particular by Melanie Krause, Roman Penslin, Rainer Schumann, and Katrin Herrmann. We kindly acknowledge helpful discussions with MBI theoretician Serguei Patchkovskii.

\subsection*{Funding}
The project was funded by Leibniz-Gemeinschaft via Grant No. SAW/2017/MBI4. Further funding is acknowledged from the Swiss National Science Foundation via Grants No. 10003697 and 193642, from DFG via Grants No. MO 719/13-1 and /14-2, from BMBF via Grant No. 05K13KT2 and from the European Union's Horizon 2020 research and innovation program under the Marie Sk\l{}odowska Curie grant agreement No. 641789. T.F., B.K., and T.v.S. acknowledge financial support by the DFG via CRC 1477 "Light-Matter Interactions at Interfaces" (ID: 441234705) and via IRTG 2676 "Imaging Quantum Systems" (ID:  437567992).

\subsection*{Author contributions}
D.R., M.V. and T.F. performed the feasibility studies in advance of the experiment. Setting up the delay stage and the microfocusing optics was done by N.M., K.K., J.Z., D.R., B.Sc., and A.R. The helium jet was set up by B.L.. D.R., J.Z., K.K., B.L., A.U. and M.S. set up the CDI detection system, and J.Z. developed the data acquisition system. N.M., B.Sc., and A.R. operated the HHG source, and J.Z., D.R., K.K., B.L., N.M., B.Sc., B.Se., M.S., R.T., A.U. and A.R. assembled and carried out the experiment. The conceptual idea of the simulation approach was conceived by T.F., M.V. and D.R. M.V., T.F., and T.v.S carried out the TDSE calculations on atomic He, and J.Z., B.K., T.F., and T.v.S. developed and performed the subsequent scattering simulations. J.Z., D.R., M.V., T.M., and T.F. analyzed the data with input from all authors. The figures were created by J.Z.,D.R.,M.S. and T.v.S. The manuscript was written and discussed with input from all authors.

The authors declare no competing financial or non-financial interests.

\subsection*{Data Availability}
The original data of this experiment is available online\\ (\url{https://share.phys.ethz.ch/~nux/paperdata/Schafer-Zimmermann2026Laser/}).

\section*{Supplementary Materials}
 
Supplementary Notes 1 to 9.\\
Supplementary Figures 1 to 12.

\printbibliography

\end{document}